\documentstyle[12pt]{article}
\newfont{\feff}{cmti10}
\def\undertext#1{\vtop{\hbox{#1}\kern 1pt \hrule}}

\def\tr{\hbox{tr}\,}

\def\be{\begin{equation}}
\def\ee{\end{equation}}
\def\bea{\begin{eqnarray}}
\def\eea{\end{eqnarray}}
\def\eqref#1{(\ref{#1})}

\begin{document}

\begin{titlepage}
\title{
\begin{flushright}
{\bf\normalsize   PUPT-1323,Revised}\\
\end{flushright}
Exact Solution of Induced  Lattice Gauge Theory at Large $N$
}

\author{
A.A. Migdal\\
Physics Department,\\
Jadwin Hall, Princeton University\\
Princeton, NJ 08544-1000\\
}
\date{June 1992}

\maketitle

\begin{abstract}

We find the exact solution of a recently proposed model of
the lattice gauge theory induced by heavy scalar field in adjoint
representation at $ N= \infty $ for arbitrary  dimension $D$. The
nonlinear integral equation for the gauge invariant density of
eigenvalues of the vacuum average of the scalar field is derived. In
the continuum limit, the density grows as $ \phi^{\alpha} $ where $
\alpha = 1 +  \frac{1}{\pi}\arccos\frac{D}{3D-2} $.

\end{abstract}
\end{titlepage}

\newpage

\section{Introduction}

Recently, the new lattice gauge model was suggested by Kazakov and
myself\cite{KazMig}. We observed, that the Yang-Mills term in the
Action could be induced by massive scalar field in adjoint
representation, in the same spirit as actual QCD is induced by some
yet unknown quark and gluon constituents. The origin and properties of
these constituents is irrelevant for the purposes of QCD, in virtue of
renormalizability.

The basic property of induced gauge models is the scaling law for the physical
mass scale $ M $ as a function of the constituent mass $ m_0 $
\begin{equation}
M \propto (m_0 - m_c)^{\gamma }
\label{Scaling}
\end{equation}
where $ m_c$ is some critical value, depending upon
the regularization scheme, and the critical index $ \gamma   $
depends upon the universal characteristics of the constituent field,
such as spin and number of components.

This scaling law comes about when one follows the running gauge
coupling constant $ \beta(r) = \frac{1}{g^2(r)} $ from the lattice
spacing scales $ r=r_0 $ where we assume $ \beta(r_0) =0 $ all the way
to the hadronic distances $ r_2 \sim \frac{1}{M} $ , where again $
\beta(r_2) \sim 0 $ due to confinement. Along the way, $ \beta(r) $
first goes up as $ c_1\ln \frac{r}{r_0} $ due to screening by the
constituent field, until we reach the scales $ r_1$ such that $ r_1^2
(m_0^2 - m_c^2)\sim 1 $ where the screening turns off, leaving us with
some large constant $ \beta(r_1) = c_1\ln \frac{r_1}{r_0} $. This
constant serves as the bare coupling for the gauge theory, with
effective ultraviolet cutoff at $ r_1$. The rest of the story is well
known: the $ \beta(r) $ goes down as $\beta(r_1) - c_2 \ln
\frac{r}{r_1} $ due to famous antiscreening from induced
selfinteraction of the gauge field, all the way to the physical scale
$ r_2$.  Comparing two parts of this $ \Lambda $ shaped curve in the
$ \ln r , \beta $ plane we find $ c_1\ln \frac{r_1}{r_0} = c_2 \ln
\frac{r_2}{r_1} $ which yields the scaling law (\ref{Scaling}).

The above arguments provide only  a rough  estimate of induced
coupling.  There are feedback effects
from the hard gluons, as well, as some induced selfinteraction of the
inducing field. If we go back to the cutoff scales from the scale $
r_1$ with some unknown constant $ \beta_1 $ we would have to compute
the renormalization of the effective $ \phi^4 $ theory coupled to
gauge field. The peculiarity of this theory is that the scalar
particle is very heavy, not like the ordinary Coleman-Weinberg
model. Clearly, the effective  $ \phi^4 $ coupling cannot be small,
otherwise the screening of charge from  scalar particle would not
overcome the antiscreening  from the gluons. Indeed, as we shall see
in that last Section, it must be
adjusted to some critical value, in order to get the scaling law.

We already know the example of the heavy particle theory which in a
large $ N $ limit becomes the string theory rather than particle
theory. This is the same scalar field in adjoint representation of $
SU(N) $ but without the gauge field. If we take the large $ N $
limit, then we are left with planar graphs, which have singularity at
some value of the coupling constant. So, if we adjust the scalar
coupling to approach this large $N$ singularity, instead of adjusting
the bare mass to get the small physical mass, then the planar graphs
are known to condense to a fishnet with lattice scale cells, leading
us to the string theory.

The problem with the latter model is well known, it does not seem to
have a continuum limit at $ D>1$. The model we are considering now,
is slightly different. If you look at the strong coupling expansion
of the induced gauge theory, by integrating the scalars first, and
representing the effective gauge action as sum over the scalar
loops\cite{KazMig}, you find the sum over closed surfaces made of
these loops glued together by sides. The typical size of the loop is
$ r_1 $, in above notations. In the large $N$ limit we expect only the
planar surfaces to survive, and produce some kind of string theory,
with possible extra  internal degrees of freedom at the world sheet.

The resulting theory must be universal, as we
expect it to fall into one of the fixed points. In four dimensions
there is only one nontrivial fixed point known, the QCD. So, the
crucial question is whether the solution of particular induced gauge
model is nontrivial. This question is a dynamical one, and could be
answered only  by exact solution.

Our best bet here is the scalar model in adjoint representation.
The unique property of this model is the possibility to diagonalize
the scalar field by the gauge
transformation
\begin{equation}
\Phi(x) = \Omega(x) \phi(x) \Omega^{\dagger}(x) \, \\;
\Omega^{\dagger}(x)\Omega(x)=1 
\end{equation}
so that the number of scalar degrees of freedom drops down from $
N^2-1 $ components of the traceless Hermitean matrix $ \Phi(x) $ to
its $ N-1 $ eigenvalues $ \phi(x) = diag(\phi_a(x)) , a=1,\dots N ,
\sum \phi_a =0$. These gauge invariant eigenvalues serve as a master
field: they freeze at a certain vacuum average, with fluctuations $
\delta \phi \sim \frac{1}{N} $.

The active degrees of freedom in the large $ N $ limit are given by
the gauge fields $ U_{\mu}(x) $, associated with the links of the
lattice. The analysis of \cite{KazMig} shows that the usual scalar
particle loops in external gauge field within the strong coupling (
random walk) expansion can be reproduced by the $ U(N) $ group
integrations over the gauge field for the frozen master field. The
summations over eigenvalues, distributed in the strong coupling
expansion according to the Wigner semi circle law
\begin{equation}
\rho(\phi) \propto  \sqrt{a^2- \phi^2}
\end{equation}
yields in the large $N$ limit the same results as Gaussian
integrations over matrix field $ \Phi(x) $.

If we accept this scenario, then the model can be dramatically
simplified by integration over gauge fields independently at each
link. The catch is that this integral \cite{IZ}
\begin{equation}
I(\Phi,\Psi) = \int D U \exp \left( N \, \tr \Phi U \Psi U^{\dagger}
\right) \propto \frac{\det_{ij} \exp(N \phi_i \psi_j )}{\Delta(\phi)
\Delta(\psi)} \label{IZ}
\end{equation}
where
\begin{equation}
\Delta(\phi) = \prod_{i<j} (\phi_i-\phi_j) \label{Vandermond}
\end{equation}
is the Vandermonde determinant, becomes at $ N \rightarrow \infty $
nontrivial statistical system in external field, given by eigenvalue
distribution
\begin{equation}
\rho_{\Phi}(\mu) = \frac{1}{N} \sum_{a} \delta(\mu - \phi_a)\\;\;
\rho_{\Psi}(\nu) = \frac{1}{N} \sum_{a} \delta(\nu - \psi_a) 
\end{equation}
This external field problem should be solved together with the saddle
point equation
\begin{equation}
\wp \int d \mu \frac{\rho_{\Phi}(\mu)}{\phi_a -\mu} = \frac{1}{2}  U'(\phi_a) -
D
\lim_{N \rightarrow \infty} \frac{1}{N}\left[\frac{\partial \ln
I(\Phi,\Psi)}{\partial \phi_a}\right]_{\Psi=\Phi} 
\end{equation}
where $ D = 4 $ is dimension of space, which we shall keep arbitrary
in the rest of this paper. We introduced the general scalar particle potential
\begin{equation}
   U(\phi) = \frac{1}{2} m_0^2 \phi^2 + \frac{1}{4} \lambda \phi^4 + \dots
\end{equation}
rather than the pure mass term considered in \cite{KazMig}. The
higher terms would be needed to get to the correct fixed point of
our model. Note, that we defined the mass term with traditional $
\frac{1}{2} $ factor, which was skipped in \cite{KazMig}.

In the previous paper \cite{KazMig} existence of two phases of this
system was conjectured. The so called weak coupling phase, with a gap
in the eigenvalue distribution reduces to the two matrix model
\begin{equation}
I(\Phi,\Psi) \propto \int d^N x d^N y \Delta(x)\Delta(y) \exp \left( N
\sum_i x_i y_i - V_1(x_i) -V_2(y_i) \right) \label{TwoMatrix}
\end{equation}
with potentials
\begin{equation}
V_{\Phi}(x) = \int d\mu \rho_{\Phi}(\mu)\ln (\mu-x) \\;\; V_{\Psi}(y)
= \int d\nu \rho_{\Psi}(\nu)\ln (\nu-y) \label{Potentials}
\end{equation}
which model was solved exactly at $ N = \infty$.

In the strong coupling phase, where there is no gap, the same
representation (\ref{TwoMatrix}) is formally valid, but potentials are
singular, which leads to drastic changes in the large $ N $ limit. The
index sums cannot be replaced by the integrals in equations of the
orthogonal polynomial method, but the individual terms become singular
and produce the necessary factor of $N$. We managed to generate
several terms of strong coupling expansion from the two matrix model,
but could not find the exact solution.

This is a pity, since the  induced QCD scenario
corresponds to the strong coupling phase, where the distribution of
eigenvalues is qualitatively the same as the semi-circle law. There
are also some internal inconsistencies in the weak coupling
solution\cite{KazMig}, in particular, there are no physical
excitations in the vacuum, as the hopping term in the effective
Lagrangean for the vacuum fluctuations vanishes in the large $N$
limit.

In the present paper we find the relevant strong coupling solution by
completely a different method.
The solution of the lattice model is
described in Section 2, where we derive the lattice version
of the equation for the density of eigenvalues of the scalar field.
The critical phenomena are
investigated in Section  3, where we find a remarkably simple local
version of this equation and solve it at the critical point.

\section{Exact Solution in the Large $N$ Limit}

We are going to apply to the one link integral the good old
Schwinger-Dyson equations. The traces of properly normalized
derivatives with respect to $ \Phi$ matrix
\begin{equation}
\left(\nabla_{\Phi} \right)_{i j} \equiv \frac{1}{N}
\frac{\partial}{\partial \Phi_{j i} } 
\end{equation}
act as traces of $ \Psi $ matrix on our integral
\begin{equation}
\tr \left(\nabla_{\Phi}^n -\Psi^n \right) I(\Phi,\Psi) =0 
\end{equation}
as it follows from the unitarity condition. Let us introduce the
matrix function
\begin{equation}
F(\Phi) = \nabla_{\Phi} \ln I(\Phi,\Psi) 
\end{equation}
and rewrite above equation as follows
\begin{equation}
\tr \left(\left(F(\Phi) + \nabla_{\Phi} \right)^n * 1 -\Psi^n \right)
=0 \label{Moments}
\end{equation}

Now, summing the geometric series, we define the matrix
\begin{equation}
 G_{\lambda}(\Phi) = \left( \lambda -F(\Phi) - \nabla_{\Phi}
\right)^{-1} * 1 
\end{equation}
which satisfies the linear differential equation
\begin{equation}
\nabla_{\Phi} G_{\lambda}(\Phi)= -1 +(\lambda -F(\Phi))
G_{\lambda}(\Phi) 
\end{equation}
and extra condition
\begin{equation}
\tr \left( G_{\lambda}(\Phi) + \frac{1}{\Psi-\lambda} \right) =0
\label{ExtraCond}
\end{equation}
The differential equation allows us to express $ G $ in terms of $ F
$ at fixed spectral parameter $ \lambda $, after which the last
condition would provide equation for $F$. One may readily check that
these equations work in the strong coupling phase, where the
function $ F(\phi) $ expands in power series of its matrix argument.
This expansion goes along with the expansion of $ G_{ \lambda}(\Phi)
$ in inverse powers of $ \lambda $, generating the moments relations
(\ref{Moments}). The expansion coefficients of $ F(\Phi) $ are
obtained one after another from the moments relations. We checked that
the same coefficients arise in the orthogonal polynomial treatment of
the two matrix model in the strong coupling phase. The first several
terms of this expansion were presented in ref\cite {KazMig}.

Now, let us take the large $ N $ limit. The key observation is that at
large $ N $ one does not have to differentiate the traces, as that
would not produce the necessary factor of $ N$. Such factors are only
produced by derivatives of the powers of the $ \Phi $ matrix
\begin{equation}
\nabla_{\Phi} \Phi^n = \sum_{k=1}^{n} \Phi^{n-k} \frac{1}{N} \tr
\Phi^{k-1} 
\end{equation}
which terms remain finite in the large $ N $ limit, when this formula
can be written in an integral form
\begin{equation}
\nabla_{\Phi} \Phi^n = \int d\mu \rho_{\Phi}(\mu) \frac{\mu^n -
\Phi^n}{\mu - \Phi} 
\end{equation}

This representation allows us to write the linear integral equation
for function $ G_{\lambda}(\nu) $
\begin{equation}
\int d \mu \rho_{\Phi}(\mu)
\frac{G_{\lambda}(\mu)-G_{\lambda}(\nu)}{\mu- \nu} = -1 + (\lambda
-F(\nu)) G_{\lambda}(\nu) \label{Gequation}
\end{equation}
As for the extra condition (\ref{ExtraCond}), it can be rewritten as
follows
\begin{equation}
\int d\mu \rho_{\Phi}(\mu) G_{\lambda}(\mu) =  \int d\nu
\frac{\rho_{\Psi}(\nu)}{\lambda -\nu}=  V_{\Psi}'(\lambda)
\label{Extra}
\end{equation}
At the moment we are interested in the spatially homogeneous master
field, so that $ \Psi = \Phi $, in which case we can omit the
subscripts of the densities and potentials.

We define these spectral
integrals in the principal value sense, so that both $ F(\mu) $ and $
G_{\lambda}(\mu) $ are real functions, when $ \lambda \, , \, \mu $ vary at
the real axis.  At $ \lambda \rightarrow \infty $ by
construction $ G_{\lambda}(\mu) \rightarrow  \frac{1}{ \lambda } $.
We shall assume that the eigenvalues are distributed along the whole
real axis, this assumption  will be justified later.\footnote{This
assumption, if fact will be valid only in the local limit, when the
small vicinity of the origin in the eigenvalue distribution will be
rescaled to the whole complex plane. In any particular lattice model,
there might be finite support, but these effects are lattice
artifacts, much in the same way, as the lattice momenta are
distributed within the cell of inverse lattice. The eigenvalues have
dimension of $ m^{\frac{1}{2} D -1} $, which means that for $ D>2 $
the cutoff for the eigenvalues grows as the power of the momentum
cutoff. In the local limit both cutoffs disappear.}

The saddle point equation allows us to eliminate one of the unknown
functions, by expressing $ F(\nu) $ in terms of the potential
\begin{equation}
F(\nu) = \frac{\frac{1}{2} U'(\nu) - \Re V'(\nu)}{D } 
\end{equation}

When this expression is substituted back into (\ref{Gequation}), the
terms combine nicely and we find the following singular integral equation
\begin{equation}
1+\wp \int d\mu \frac{\rho(\mu)G_{\lambda}(\mu)}{\mu- \nu} =
(\lambda-R(\nu)) G_{\lambda}(\nu) \\; \; R(\nu) = F(\nu)+\Re V'(\nu)
\end{equation}
This equation could be regarded as  the boundary value problem for
analytic function
\begin{equation}
{\cal T}_{\lambda}(z) = 1 + \int d\mu
\frac{\rho(\mu)G_{\lambda}(\mu)}{\mu-z}
\end{equation}
in the upper half of complex $ z $ plane, with usual analytic
continuation to the whole plane by symmetry
\begin{equation}
  {\cal T}_{\lambda}(\bar{z}) = \bar{{\cal T}}_{\lambda}(z)
\end{equation}
In virtue of symmetry of the eigenvalue distribution we could have
mapped the upper half plane to the cut plane by conformal transformation $
\zeta=-z^2 $ but it appears to be simpler to work in the upper
half plane, keeping in mind the symmetry of boundary values at the real
axis.

We shall assume the convergent integrals, in which case
\begin{equation}
{\cal T}_{\lambda}(\infty) = 1
\end{equation}
The next term of asymptotic expansion at $ z \rightarrow\infty$ is
also known in virtue of an extra condition (\ref{Extra}), namely
\begin{equation}
{\cal T}_{\lambda}(z) \rightarrow 1 - \frac{\Re V'(\lambda)}{z}
\label{Asymptotics}
\end{equation}

The boundary problems of this kind can be solved in general.  We
express $ G $ in terms of discontinuity
\begin{equation}
\Im {\cal T}_{\lambda}(\nu_+)= \pi \rho(\mu) G_{\lambda}(\mu)
\label{DiscF}
\end{equation}
and find the following ratio of the complex conjugate  boundary
values of ${\cal T}$ at $ \nu_{\pm} $
\begin{equation}
\frac{{\cal T}_{\lambda}(\nu_+)}{{\cal T}_{\lambda}(\nu_-)} =
\frac{\lambda-R(\nu)+ \imath \pi \rho(\nu)}{\lambda-R(\nu)-\imath \pi
\rho(\nu)}
\label{Bproblem}
\end{equation}

The boundary problem has a unique solution
\begin{equation}
{\cal T}_{\lambda}(z) = \exp \left( \int_{- \infty}^{\infty} \frac{d \nu}{ \pi
(\nu-
z )} \arctan \frac{\pi \rho(\nu)}{\lambda-R(\nu)}\right)
\label{Solution}
\end{equation}
The ambiguity of the choice of the branch of arctangent is removed by
an extra requirement that at $ \lambda \rightarrow \infty$
\begin{equation}
{\cal T}_{\lambda}(z) \rightarrow 1 -\frac{V'(z)}{\lambda}
\end{equation}

Comparing the first term at $ z \rightarrow \infty$ with the
asymptotics (\ref{Asymptotics}) we find the following integral equation
for the potential
\begin{eqnarray}
& &\Re V'(\lambda)= \int_{- \infty}^{+ \infty} \frac{d \nu}{ \pi} \,
\arctan \frac{\pi \rho(\nu)}{\lambda-R(\nu)} \\ \nonumber & \,&
\Im V'(\nu_+) = -\pi\rho(\nu) \\ \nonumber & \,&
R(\nu) = \frac{1}{2 D}U'(\nu) + \frac{D-1}{D}\Re V'(\nu)
\label{Master}
\end{eqnarray}

We shall refer to this equation as  the Master Field Equation, or
MFE. In
Appendix we study general properties of this equation, assuming finite
support.

\section{Critical Point}

Let us  study the small vicinity of the origin in the MFE. The idea
is, that the eigenvalues scale as positive power of mass, therefore,
in the lattice units we are using, the continuum theory corresponds to
infinitesimal vicinity of the origin. The normalization of the density
is also the lattice artifact, as it follows from dimensional
countuing. The internal nonlinearities in the scaling region, rather
then the global normalization, would fix the scale of the density at
the origin.

With this picture in mind, let us restore the
imaginary part of MFE  from
analyticity  in the upper half plane
\begin{equation}
 V'(z) = \int_{- \infty}^{ \infty} \frac{d \nu}{\pi \imath} \,
\ln \left(1 + \frac{\imath \pi \rho(\nu)}{z-R(\nu)} \right)
\end{equation}
In the lower half plane we would have $ -\imath $ instead of $ \imath
$ in the same integral.\footnote{Let us stress once again, that this formula is
not valid for finite support of the eigenvalues. The correct formulas
in this case are studied in Appendix, and this one can be obtained
from the general formulas in the limit $ a \rightarrow \infty$.}

To make it less surprizing, let us consider $ \Omega = \frac{V'(z)}{z}
$ as analytic function of $ \zeta= -z^2 $, which it is, since the
distribution of eigenvalues is symmetric.  Then the upper half plane
would map onto the first sheet of the Riemann surface, the real $ z $
axis being mapped onto the cut from $ - \infty $ to $ 0$ in the $
\zeta $ plane, with negative(positive) $ z $ corresponding to
upper(lower) side.

Imaginary part of $ V'(z_+) $ is negative both for
positive and negative $ z $, and the real part changes sign with $ z
$, therefore the values of $ \Omega(\zeta) $ at the opposite sides of
the cut are complex conjugate, as usual. The second sheet function,
with $ \imath \rightarrow - \imath $ has the opposite sign of
the imaginary part, so it matches the corresponding boundary values
at the cut, as it should.

We went into these boring details, because
it took us some time to turn away from the usual picture with finite cut
in the $ z $ plane and no singularity at infinity. We are going to
use the old $ z $ variable in the rest of the paper, but restrict it
to the upper half plane.

Let us study the equation for the imaginary part $- \pi \rho(\lambda)
$ of $ V'(\lambda_+) $. The imaginary part of MFE reads
\begin{equation}
 \pi \rho(\lambda) = \int_{- \infty}^{ \infty} \frac{d \nu}{2\pi} \,
\ln \left( 1 +
\frac{\pi^2 \rho^2(\nu)}{(R(\nu) - \lambda)^2} \right) .
\label{MFCont}
\end{equation}

Let us differentiate this equation, we find
\begin{equation}
 \rho'(\lambda) = - \wp \int_{- \infty}^{ \infty} d \nu \,
\frac{\rho^2(\nu)}
{(\lambda-R(\nu))\left((\lambda-R(\nu))^2 + \pi^2 \rho^2(\nu)\right)}
\label{RHO'}
\end{equation}
Now, investigation of this equation reveals that critical phenomena
take place when the renormalized scalar potential
\begin{equation}
 2 u(\phi) = U(\phi) -  D \phi^2 = \frac{1}{2} m^2 \phi^2 + \frac{1}{4} g
\phi^4
\end{equation}
goes to zero in the lattice units we are using, so that $ R(\nu)
\rightarrow \nu $.\footnote{The
remarkable fact, that in spite of all the strong fluctuations of the
link variables, which were very far from continuum, we still get the
same critical value $ m_c^2 = 2D $ of the bare mass, as in
asymptotically free gauge theory, would be discussed later.}

The relevant eigenvalues would  scale as some power of scalar mass $
\nu \sim m^{\gamma} $, and density  would scale as $ \rho \sim m^{2+
\gamma}$, the same as $ u'(\phi) $ where the critical index $ \gamma
$ will be computed below.

The scaling solution comes about in a tricky way. There are two relevant
regions of integration in (\ref{RHO'}). In the region  of $ |\nu - \lambda|
\sim
m^{\gamma} $,  we could neglect   $ \rho $ in the denominator, which
yields the standard principle value integral. Note, that the size of
the gap $  \delta \nu $ which is needed to define the principle value
integral, could be any number in the interval $ m^{2 + \gamma} \ll
\delta \nu \ll m^{\gamma} $, the result would not depend upon this gap then.

There is another, much
smaller region,
that of $ |\nu - \lambda| \sim m^{\gamma+2} \ll \delta \nu  $ where we cannot
neglect $ \rho $ but we can expand $ u'(\nu) $ and $ \rho(\nu) $ in
Taylor series near $ \nu = \lambda $ after which integration reduces
to the residue in complex pole. The integral converges in this region,
so that the boundary $ \delta \nu $ between regions drops from the answer.

Collecting the terms, we find,
that the term $ \rho'(\lambda ) $  cancels, so that we are
left with the  following integrodifferential equation
\begin{eqnarray}
& & \left(\rho(\lambda )r(\lambda )\right)''  = \mbox{regular terms}-  \wp
\int_{- \infty}^{ \infty} d\nu \,
\frac{\rho^2(\nu)}{(\lambda -\nu)^3} \\
\nonumber & \,&
r(\lambda ) = u'(\lambda )+\frac{D-1}{D}
\wp \int_{- \infty}^{ \infty} d\mu \,\frac{\rho(\mu)}{\lambda -\mu}
\label{scaling}
\end{eqnarray}
The regular terms represent the  contributions from the
cutoff scales. As usual in the field theory, such contributions provide
the contact terms, without singularities. Note, that this equation
could be integrated twice in $ \nu $ at the expence of redefinition
of the regular terms. On the left side we would obtain
$\rho(\lambda )r(\lambda )$, and on the right the denominator $
(\lambda -\nu)^3 $
would be replaced by $ 2 (\lambda-\nu) $.

This convenient method of elimination of regular terms is to
introduce two analytic  functions
\begin{equation}
  {\cal P}(z) = \frac{D}{1-D}u'(z)-V'(z)
\end{equation}
\begin{equation}
  {\cal Q}(z) =  \mbox{regular terms}
- \pi \int_{- \infty}^{ \infty} d\mu \,\frac{\rho^2(\mu)}{z-\mu}
\end{equation}
and note that at $ \Im z \rightarrow +0$ by construction
\begin{equation}
  \Im\,{\cal Q} =  \left(\Im \,{\cal P}\right)^2.
\end{equation}
On the other hand, in virtue of the  above equation for density
\begin{equation}
  \Re \, {\cal Q} = \frac{1-D}{D} \Im \,\left({\cal P}^2\right)
\end{equation}

These two equations should be supplied with the symmetry conditions,
namely, imaginary parts of these two functions at  $ \Im z \rightarrow
+0$ must be even functions of $ z $, whereas real parts must be odd.
The derivative of the real part of $ {\cal P}(z)$  at the origin
plays the role of effective scalar mass $ m^2 $.

This nonlinear boundary problem, which we call Local Master Field
Equation, (LMFE) represents the main result of this work. At first
glance it looks too simple to have nontrivial solution, but this is
not true, as we found  by analytical and numerical studies.

We doubt that there could exist an exact solution in the general
case, but precisely at the critical point, corresponding to the
vanishing mass, such solution can be found in virtue of scale
invariance. Let us take the  powerlike Anzatz
\begin{equation}
   {\cal P}(z) = p\, z^{\alpha}\\;\;
   {\cal Q}(z) = q\, z^{2 \alpha}\\;\; \alpha = 1 +  \frac{2}{ \gamma}
\end{equation}
The symmetry comditions would be satisfied provided
\begin{equation}
  p = \imath p_0 \exp \left(-\frac{1}{2} \imath \pi \alpha \right)\\;\;
  q = \imath q_0 \exp \left(-\imath \pi \alpha \right)
\end{equation}
where $ p_0$ and $q_0$ are real numbers. Now, the powers of $ z $
would cancel in LMFE, so that we may check it at $ z=1$ where
\begin{equation}
  \Im\, {\cal P} = p_0 \cos \frac{1}{2} \pi \alpha \\;\;
  \Im\, {\cal P}^2 = p_0^2 \sin \pi \alpha \\;\;
  \Im\, {\cal Q} = q_0 \cos \pi \alpha \\;\;
  \Re\, {\cal Q} = q_0 \sin \pi \alpha
\end{equation}
Comparing this with  LMFE, we find
\begin{equation}
  \cos \pi \alpha  = -\frac{D}{3D-2} \\;\;
\cos \frac{2\pi}{\gamma}  = \frac{D}{3D-2} \\;\;
q_0 = \frac{1-D}{D} p_0^2
\end{equation}
The index $ \gamma(D) $ takes special values $ \gamma(0) = 4,\gamma(1)
= \infty, \gamma(2) = 6 $ at low dimension, after which it smoothly
goes to its asymptotic value $ \gamma(\infty) = 5 $ taking only
transcendental values at integer dimensions. In particular, $
\gamma(4) = 5.41991, \alpha(4) = 1.36901 $. Note, that unlike the
singularity at $ D=1 $ which was manifest in initial model, triviality
of the theory at $ D=2 $ comes about in a very nontrivial way. In our
opinion, this is an independent evidence for equivalence of this model
to the gauge theory.

The general scaling solution
\begin{equation}
   \rho(\phi) = m^{2+\gamma} f \left(\frac{\phi}{m^{\gamma}} \right)
\end{equation}
starts from some constant at $ \phi =0 $,
then grows, approaching the above powerlike asymptotics, $ \rho
\propto \phi^{\alpha} $ at $ \phi \gg m^{\gamma} $, so that the mass
cancels.

This is the magnified vicinity of the origin in complete spectrum of
eigenvalues; apparently, the density reaches the maximum, and then
starts decreasing outside the scaling region. The density at the
origin goes to zero, so that at negative $ m^2 $ the gap would
appear in the distribution (clearly, one cannot continue our
irrational powers of $ m $ to negative $ m^2 $ ). In the forthcoming
publication\cite{Louqi} we solve this equation numerically at various
$ D $.

As discussed in the previous paper\cite{KazMig}, the scaling laws in
this model translate into logarithmic laws of original theory: for the
effective induced gauge coupling of the form
\begin{equation}
  \frac{1}{g_0^2} \sim b N \ln \frac{\Lambda^2}{m_0^2-2D}
\end{equation}
the physical mass scale would be
\begin{equation}
  M^2 \sim (m_0^2-2D) \exp \left(-\frac{48 \pi^2}{11 N g_0^2} \right)
\sim (m_0^2 -2D) \left(\frac{m_0^2-2D}{\Lambda^2} \right)^{\frac{48 \pi^2
b}{11} }
\end{equation}

Should we identify this scale with the scale of our field
$(m_0^2-2D)^{2 \gamma} $ we would conclude that
\begin{equation}
b = \frac{11}{48 \pi^2} \,
\left(\frac{ 4\pi}{\arccos\frac{2}{5}}-1 \right)
\end{equation}
but this seems to be too long a shot: one should
investigate the physical mass spectrum before drawing any conclusions.

\section{Acknowledgements}

I am grateful to David Gross and Sasha Polyakov for fruitful
discussions and important remarks. Special thanks to Volodya Kazakov
whose visit here in April was crucial for this project. The help of
Vadim Borue in investigation of the MF equation was also very
important. I was supported by the National Science Foundation under
contract PHYS-90-21984.

\appendix

\section{Investigation of the Master Field Equation}

Let us study the Master Filed Equation (\ref{Master}) assuming the
finite support  with the
following Anzatz for the normalized spectral density
\begin{equation}
\nu = a \cos( \alpha) , \, 0< \alpha < \pi, \,\rho(\nu)= \frac{1}{\pi}
\sum_{k=1}^{ \infty} \xi_k \frac{1-\cos( 2k \alpha) }{a \sin( \alpha)}
, \, \sum_{k=1}^{ \infty} \xi_k =1 \label{Anzatz}
\end{equation}
This is just the Fourier expansion, with correct symmetry properties.
The corresponding Fourier expansion for the real part of potential can
be found by elementary integration
\begin{equation}
\Re V'(a \cos(\beta)) = \wp \int_{-a}^a d\nu \frac{\rho(\nu)}{a
\cos(\beta) -\nu}= \sum_{k=1}^{ \infty} \xi_k \frac{\sin( 2 k
\beta)}{a \sin( \beta)} 
\end{equation}
and the potential as analytic function
\begin{equation}
V'(z) = \frac{1}{\sqrt{z^2-a^2}} \left(	1-\sum_{k=1}^{ \infty} \xi_k
\left(\frac{z-\sqrt{z^2-a^2}}{a} 	\right)^{2k} \right)
\label{Potential}
\end{equation}

Substituting in (\ref{Master}) the previous formula, and Fourier
transforming, we find
\begin{equation}
\xi_k = \frac{a}{2\pi^2} \int_0^{\pi} d \alpha \,\sin \alpha \,
\oint_{-a}^{a}d \lambda \, \ln \left(\frac{\lambda- R - \imath \pi
\rho }{\lambda-R + \imath \pi\rho}\right) \,
\left(\frac{\lambda - \sqrt{\lambda^2-a^2}
}{a}\right)^{2 k} -\{\rho \rightarrow 0\} 
\end{equation}
where we skipped for brevity the arguments of $ R, \rho  $.  The contour
encircles anticlockwise the interval $ -a,a $. The last factor takes
complex conjugate values $ \exp (\pm 2\imath k \beta ) $ at $ \lambda=
a \cos \beta $, which yields correct integrand after taking
discontinuity. The term with $ \rho = 0 $ is substructed, since this
is how the branch of arctangent was defined.

The $ \lambda $ integrand has only two singularities outside the
contour, and decreases as integer negative power of $ \lambda $, so
that there is no residue at infinity. The discontinuity along the cut
from $ R- \imath \rho  $ to $ R+\imath \rho$ is equal to $ \pi $, and this
integral
is elementary. As a result we find
\begin{equation}
\xi_k = \frac{a^2}{\pi} \Im \, \int_0^{\pi} d\alpha \,\sin \alpha \,
\left(	\frac{\left(C-\sqrt{C^2-1}\right)^{2k-1}}{2k-1}-
\frac{\left(C-\sqrt{C^2-1}\right)^{2k+1}}{2k+1} \right) - \{C
\rightarrow \Re \,C  \}\label{Xiequation}
\end{equation}
where
\begin{equation}
C\equiv \frac{R(\alpha) + \imath \pi \rho(\alpha)}{a} = \frac{m_0^2}{2 D}\cos
\alpha +\frac{D-1}{ a D}\Re
V'\left(a \cos \alpha\right) -\frac{\imath}{a}\Im V'\left(a \cos
\alpha + \imath 0\right) \label{Comega}
\end{equation}
We restricted ourselves here to pure mass term, without interaction
terms in the scalar potential.

Summing up the equation for $ \xi_k $ from $ k=1 $ to $ \infty $ ,
using normalization condition (\ref{Anzatz}) on the left, and summing up the
series on the right, we find the following equation for $ a $
\begin{equation}
\frac{\pi}{a^2} = \int_0^{\pi} d \alpha \,\sin \alpha \, \Im \left(
C-\sqrt{C^2-1} + \sqrt{(\Re\,C)^2-1}\right) \label{Aequation}
\end{equation}
But the same normalization condition yields the simpler equation
\begin{equation}
\frac{\pi}{a^2} = \int_0^{\pi} d \alpha \,\sin \alpha \, \Im \,C
\end{equation}
which means that contribution from the square roots should
cancel. This relation can be written as follows
\begin{equation}
0=\oint \frac{d \omega}{2 \imath \omega} \frac{1-\omega^2}{2 \imath
\omega} \left(\sqrt{C^2-1} -\sqrt{(\Re\,C)^2-1} \right)
\\; \; \omega = e^{\imath \alpha} \label{ANorm}
\end{equation}

The explicit expression for the analytic potential (\ref{Potential})
follows from above representation
\begin{equation}
V'(z) = \frac{2}{a^2}\left(z-\sqrt{z^2-a^2}\right)-a\, {\cal U }
\left(\frac{z-\sqrt{z^2-a^2}}{a} \right)  
\end{equation}
where
\begin{equation}
{\cal U }(\zeta) = \int_{0}^{2\pi} \frac{d \alpha \sin \alpha}{2 \pi
\imath}\left( 2 \zeta \left(C - \sqrt{C^2-1}\right)-\ln \frac{(C +
\sqrt{C^2-1}+ \zeta)} 	{(C + \sqrt{C^2-1}- \zeta)} \right) - \{C
\rightarrow \Re \,C  \}
\label{FinalPhi}
\end{equation}

Now, using (\ref{Comega}) we can write down the integral equation for
$ C(\alpha) $.
\begin{equation}
C(\alpha)= \frac{a^2 m_0^2 - 4}{4a^2 D}\omega + \frac{a^2 m_0^2 + 4D-2}
{2a^2 D\omega} + \frac{2D-1}{2D}{\cal U
}\left(\frac{1}{\omega}\right) -\frac{1}{2D}{\cal U
}\left(\omega\right) \label{FinalC}
\end{equation}
which should be solved together with (\ref{ANorm}).

The semi circle law at $m_0^2 \rightarrow \infty$ corresponds to the
following
\begin{equation}
a^2 \rightarrow \frac{4}{m_0^2} \\; \; C(\alpha) \rightarrow -
\frac{m_0^2}{\omega} \label{Wigner}
\end{equation}
since ${\cal U }(\zeta) \rightarrow 0 $ as $ C^{-3} $ at large $ C $.
The function $ C(\omega) $ becomes analytic at infinity, which allows
us to expand the contour and in (\ref{ANorm}) to infinity, where
expansion goes in even inverse powers of $ \omega $, so that there is
no residue. As for the integral with the real part, here we cannot
expand the contour, but the integral is small by itself, as the
imaginary part of this square root does not vanish only in the small
region of angles near $ \frac{1}{2} \pi $.

We solved these equations  at $D=4$  numerically by iterations, taking
as unknowns the values of $ C $ at  $ M = 4096 $ of points in the first
quadrant
\begin{equation}
C_m = C\left(\alpha = \frac{\pi m}{2 M +1}\right)\\; \; 0<m\leq M \\; \; C_{2
M+1-m} = -\bar{C}_m
\end{equation} 
The integral was approximated by sum over these equidistant points,
which corresponds to the Gauss-Tchebyshev method.

As expected, at large $ m_0^2 $ we reproduced the semi circle,
moreover, it was difficult to get anything else.
The nontrivial term $ U $ in our equation was practically negligible,
unless we adjusted the bare mass so, that the derivative $ \Re V''(0)
$ vanished. Then we started seeing interesting phenomena, but in order
to compare them with our local equations, much better resolution would
be needed. According to our analysis in the Section 3, there are two
different small scales involved, and there are strong cancellations.
In such situation, the approximation of the integral by sum over equidistant
points is no longer valid. The corresponding numerical solution is
studied in the forthcoming work\cite{Louqi}

\end{document}